# Haptic Reproduction of Curved Surface and Edge by Controlling the Contact Position between the Disk and the Finger Using Airborne Ultrasound

Aoba Sekiguchi[1], Tao Morisaki[2], Takaaki Kamigaki[1], Yasutoshi Makino[1] and Hiroyuki Shinoda[1]

[1] Graduate School of Frontier Sciences, The University of Tokyo, Japan

[2] NTT Communication Science Laboratories, Nippon Telegraph and Telephone Corporation, Atsugi, Japan

(Email: sekiguchi@hapis.k.u-tokyo.ac.jp)

**Abstract ---** By presenting curved surfaces of various curvatures including edges to the fingertip, it is possible to reproduce the haptic sensation of object shapes that cannot be reproduced by flat surfaces alone, such as spheres and rectangular objects. In this paper, we propose a method of presenting curved surfaces by controlling the inclination of a disk in contact with the finger belly with acoustic radiation pressure of ultrasound. The user only needs to mount a lightweight device on the fingertip to experience a tactile presentation with low physical burden. In the demonstration, the user can experience the sensation of stroking an edge and different curvatures of curved surfaces.

**Keywords:** airborne ultrasound, passive device, curved surface

## 1 Introduction

Haptic presentation technology reproducing the sensation of touching an object is expected to be used to improve the immersive of virtual reality (VR). VR controllers for consumers already incorporate vibrators and control the timing of vibrations (e.g., by driving the vibrators in accordance with the timing of collisions between hands and objects) to enhance the immersive of content and the sense of operation.

In a VR space, if the shape of an object (sphere, rectangle, etc.) can be identified by haptic sensation alone as in reality, the immersive is further enhanced compared to simple vibration alone. Since the contact position between a fingertip and an object changes depending on the way the fingertip touches the object at that time, haptic reproduction of object shapes requires control of the spatial distribution of the haptic stimulation presented. This spatial distribution control has been attempted with pin arrays [1] and electrode arrays (arrayed electrotactile displays). Verschoor et al. have reproduced the sensation of touching various objects by controlling the angle of contact between the disk and the finger belly [2]. In this disk device, for example, the sensation of touching a flat surface is presented by pressing a flat plate against the finger belly. The sensation of touching a curved surface, such as a sphere, is presented by changing the contact position between the disk and the finger belly to match the contact position between the curved surface and the finger belly.

However, haptic devices reproducing spatial distribution tend to be complex and bulky. In the case of pin arrays, individual wiring connecting to drive circuits is required for each element. In addition, in the method using the contact angle of the disks described above, three servo motors are used to control the angle [2].

In order to reproduce haptic shapes with a lightweight and simple mechanism, we propose a disk device that is remotely driven by ultrasound (Fig:1)

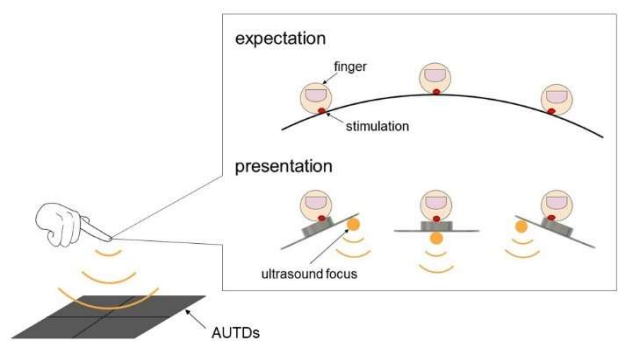

Fig1: Conceptual diagram

based on the disk device by Verschoor et al. In the proposed method, a plastic disk weighing only 1.85 g is mounted to the user's fingertip. The disk is remotely driven by focused ultrasound, and the non-contacting force (acoustic radiation pressure)[3] generated at the focal point controls the inclination of the disk, thereby changing contact position between finger and disk. In this method, when presenting the sensation of stroking a curved surface, the disk is continuously inclined in accordance with the horizontal movement of the fingertip, and the contact position between the finger and the disk is also moved horizontally [4]. By increasing the speed of movement of the contact position to an extremely high speed, it is possible to reproduce the sensation of stroking a pointy surface with a radius of curvature of 0 mm.

Although devices driven by acoustic radiation pressure of ultrasound have been proposed in the past [5], the purpose of this device is to amplify the radiation force with a leverage mechanism and present it as a strong force stimulus. This device doesn't control the spatial distribution of the force, and thus does not present curved surfaces or shapes. The disk device in this study also uses a leverage mechanism to amplify the stimulus and change the stimulus position to reproduce the sensation of stroking a sharp surface. There is a study that reproduced the sensation of stroking a curved surface by irradiating ultrasonic waves directly to the skin[6], but the stimulus was weaker than when the disk device was worn, and the proposed device did not reproduce the sensation of stroking a curved surface as the proposed device does.

## 2 METHODS

In this chapter, we introduce the developed ultrasound-driven haptic disk device and its driving system. We also describe the algorithm for driving the device to reproduce the haptic sensation of stroking curved surfaces and edges.

### 2.1 Device

Fig2 shows a photograph and a schematic diagram of the developed haptic disk device. This disk device consists of a 15 mm diameter disk (disk for stimulating)

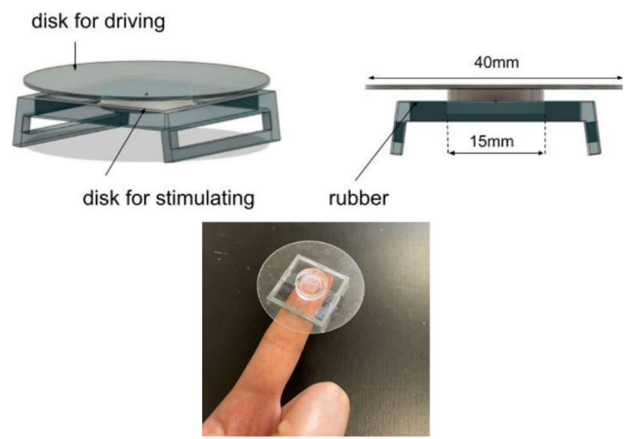

Fig2: Haptic disk device driven by ultrasound to contact the skin and present tactile sensations, a 40 mm diameter disk (dis for driving) to irradiate ultrasound focus and drive the device, and a mounting part to mount it to the finger. The disks are made of acrylic, and the mounting part is made of resin (Formlabs Gray Resin V4) with a Young's modulus of 2.77 GPa using an optical 3D printer (Form3+, Formlabs). The disk for stimulating, the disk for driving, and the mounting part are connected by a rubber membrane, and the elasticity of this connection allows the disk to change its inclination in response to the force applied to the disk for driving.

### 2.2 Driving system

Fig3 shows the driving system for the haptic disk device. The system consists of a depth camera (Intel RealSense Depth Camera D435) for tracking the 3D position of the disk device and Airborne Ultrasound Tactile Displays for generating an ultrasound focus and driving the disk device [7]. A red color marker is pre-positioned at the center of the disk for driving, and the device position is tracked by detecting the position of the marker with a binary color filter. An Airborne Ultrasound Tactile Display is an array of ultrasound transducers whose phases can be individually controlled. By controlling the phase appropriately, an ultrasound focus can be generated at an arbitrary location in space, and a force is applied remotely by presenting this focus to the disk device. In this paper, 996 ultrasound transducers (40 kHz) are used.

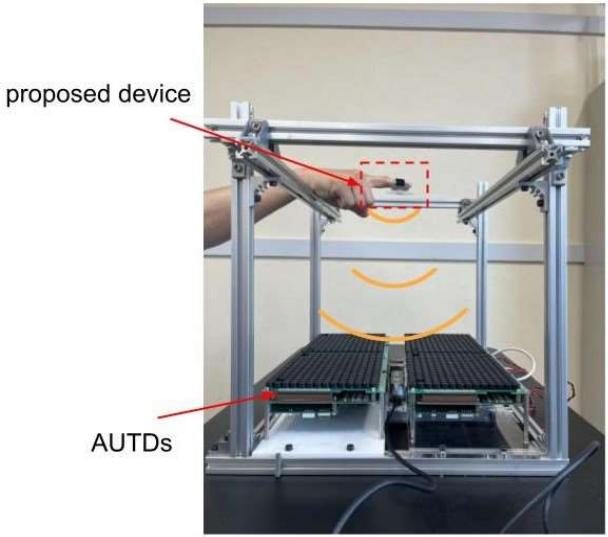

Fig3:Setup of haptic presentation

### 2.3 Driving algorithms for curved surface presentation

Fig4 shows a schematic diagram of the driving algorithm (algorithm for determining the focal point) of the haptic disk device to reproduce the sensation of stroking a curved surface. The ultrasound focus is moved in accordance with the finger movement, and the contact position between the disk and the finger is changed to present a curved surface. We assume that the fingers move in one direction. We define $2r_{fo}$ as the maximum focus movement distance on the disk (within the diameter of the disk), $x_{fo}$ as the x-coordinate of the focus on the disk, $2r_{fin}$ as the maximum movement distance of the finger, and $x_{fin}$ as the x-coordinate of the finger (Fig4). The focus shift distance on the disk $x_{fo}$ is moved according to the following equation.

$$x_{fo} = -\alpha \frac{r_{fo}}{r_{fin}} x_{fin} \quad\quad 1$$

$\alpha$ is a coefficient that determines the speed of focus movement relative to finger movement, and its range is $0 < \alpha \leq 1$. When $\alpha = 1$, the focus moves fastest, and the focus moves from one end of the disk to the other ($r_{fo}$ to $-r_{fo}$) until the user moves his finger from $-r_{fin}$ to $r_{fin}$.

Unlike the expression (Formula 1), the sensation of stroking the edge is presented by changing the focal point in two steps according to the position of the finger. In this case, $x_{fo}$ is as follows.

$$x_{fo} = \begin{cases} r_{fo}, & x_{fin} \leq 0 \\ -r_{fo}, & x_{fin} > 0 \end{cases} \quad\quad 2$$

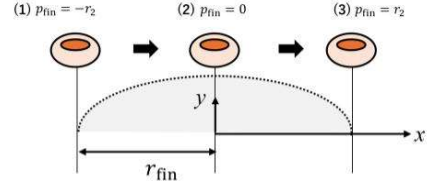

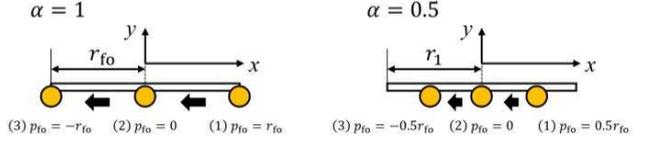

Fig4: Pattern of focus irradiation on a disk to reproduce the haptic sensation of stroking a curved surface. The focus position is moved horizontally (in the x direction) based on the positional relationship between the fingertip and the assumed curved surface.

## 3 CONCLUSION

In this paper, we reproduce the sensation of stroking curved surfaces and edges by remotely controlling the contact position between the disk and the finger using ultrasound radiation pressure. In the demonstration, the user mounts the device described in section 2.1 to reproduce the sensation of stroking curved surfaces and edges.